\documentstyle{article}

\def\p#1#2{{\partial{#1}\over\partial{#2}}}

\def\plni#1#2#3{\left(\partial\ln#1\over\partial\ln#2\right)_{#3}}

\begin{document}

\title{Application of the Galerkin method to the
problem of stellar stability, gravitational collapse and black hole formation}

\author{G.S.Bisnovatyi-Kogan\thanks{IKI RAN, Email: gkogan@mx.iki.rssi.ru}
 \and A.V.Dorodnitsyn\thanks{IKI RAN, Email: dora@mx.iki.rssi.ru}}

 \maketitle

\begin{abstract}
Approximate approach is suggested for investigation of
equilibrium stellar models, a relativistic collapse problem and black
hole formation, based on a Galerkin method. Some results of its simplified
version - energetic method- are reviewed, and equation for general
Gelerkin method are presented.
\end{abstract}

\section{Introduction}
Relativistic collapse leading to a black hole formation
is a final stage of the evolution of the sufficiently massive stars
with a mass $M > (30\,-\,50)M_{\odot}$. To describe these collapse
dynamical equations in general relativity (GR) are solved numerically
(Baumgarte et al., 1995, 1996). In order to simplify the problem for obtaining
a rapid approximate answer for such a questions, like collapse behaviour
of different stellar masses, comparative neutrino light curves and their
spectra, a simplified method is suggested. It is based on the classical
Galerkin method, widely used in different mechanical and physical
problems, connected with numerical computations (Fletcher, 1984).
For the problem of stellar collapse this method may be considered as a
generalization of the well known energetic method, which was applied
for calculations of stellar stability (Zeldovich and Novikov, 1965, 1967) and
initial stages of stellar collapse (Bisnovatyi-Kogan, 1968)
in post-newtonian approximation (PN). Here we formulate the Galerkin method
for the problems of relativistic collapse in GR.

\section{Energetic method in post-newtonian approximation (PN)}

Consider first the formulation of the energetic method in PN. The main idea
of this method is an averaging of the static or dynamic equations over the
volume, using a given prescribed density distribution inside the star.
We consider here only spherically symmetric stars, where the
density is prescribed by polytropic distribution with a
central density, obtained from algebraic (static) or ordinary
differential (dynamics) equation.
Consider polytropic equation of state $P=K\rho^\gamma$.
Combining the equilibrium and continuity equations,
we get the equation for $\rho(r)$
\begin{equation}
\label{ref34.1}
  {d\over dr}\left[K\gamma r^2\rho^{\gamma-2}{d\rho\over dr}\right]
  =-4\pi G\rho r^2.
\end{equation}
Where $r$ is the Euler radial coordinate, $\rho$ is the density.
Transforming to dimensionless quantities ( Chandrasekhar, 1939 ) $\theta$ and $\xi$,
defined as
\begin{equation}
\label{ref34.2}
  \rho=\rho_c\theta^n,\quad r=\alpha\xi,\quad
  \alpha=\left[{(n+1)K\over 4\pi G}\rho_c^{{1\over n}-1}\right]^{1/2}
  \gamma=1+{1\over n}\mbox{,}
\end{equation}
where $n$ is an polytrope index, $\rho_c$ is the central density,
 gives the equilibrium Lane-Emden equation
\begin{equation}
  \label{ref34.3}
  {1\over \xi^2}\>{d\over d\xi}\left(\xi^2 {d\theta\over d\xi}\right)
  =-\theta^n
\end{equation}
 with the boundary conditions
  $\theta=1,\quad {d\theta\over d\xi}=0~~\hbox{at $\xi=0$}$.
The stellar boundary corresponds to $\xi=\xi_1$ so that $\theta(\xi_1)=0$.
The stellar mass $M$ expressed in terms of variables (\ref{ref34.2}) becomes
 $$ M=4\pi\int_0^R \rho r^2 dr=4\pi\rho_c\alpha^3\int_0^{{\xi_1}}
  \theta^n\xi^2 d\xi $$
\begin{equation}
\label{ref34.5}
  =4\pi\left[{(n+1)K\over 4\pi G}\right]^{3/2}
  \rho_c^{{3\over 2n}-{1\over 2}}\int_0^{{\xi_1}} \theta^n\xi^2 d\xi.
\end{equation}
 and hence from (\ref{ref34.3}),
\begin{equation}
  \label{ref34.6}
  \int_0^{\xi} \theta^n\xi^2 d\xi=-\xi^2{d\theta\over d\xi},\quad
  -\xi^2{d\theta\over d\xi}\Bigg|_{\xi=\xi_1}=M_n.
\end{equation}
Obviously, at $n=3$, $\gamma=4/3$ the stellar mass is independent of
$\rho_c$ and is exactly determined by the constant $K$ in the equation of
state. The stellar mass increases at $\gamma>4/3$
with increasing $\rho_c$, and at $\gamma<4/3$
decreases. If the polytropic power coincides with the adiabatic power,
$\gamma=\gamma_{\rm {ad}}$, then the star is stable at $\gamma>4/3$  and
unstable at $\gamma<4/3$, and $\gamma=4/3$ corresponds to the boundary case
and represents the indifferent equilibrium.

Real stars are not polytropes, but the condition $\gamma=4/3$ is
approximately valid at the boundary of stability if $\gamma$ is treated
as an adiabatic power properly averaged over the star.
A strict derivation of stability conditions, including GR, is made by use
of the variational method (Harrison et al., 1965).
Equating the first variation to zero yields
the equilibrium equation, while the stability condition requires the second
variation to be positive. For an isentropic polytrope with $\gamma=4/3$,
$\rho_c$ is arbitrary, whereas the density distribution $\theta(\xi)$
is invariant against homologous contraction or expansion. Let us treat these
properties as valid also for the case where is $\gamma=4/3$ only in average.
We then derive the equilibrium and stability conditions, using the
simplified variational method based on the assumption of homology and
conservation of stellar structure at density variations
(Zeldovich and Novikov, 1967), usually
called energetic method. Write down the total energy of an instantaneously
static star analogous to the potential energy of consevative mechanical
system:
\begin{equation}
\label{ref34.7}
  \varepsilon=\int_0^M E(\rho,T)\,dm-\int_0^M {Gm\,dm\over r}
  -5.06\,{G^2M^3\over R^2c^2},\quad
  dm=4\pi\rho r^2dr
\end{equation}
 The first term here represents the internal energy $\varepsilon_{\rm  i}$,
the second the newtonian gravitational energy $\varepsilon_G$, and the third,
$\varepsilon_{\rm {GR}}$, a small GR correction
$\left({r_g\over r}={2Gm\over c^2r}\ll 1~\hbox{is the small parameter}
\right)$ evaluated by Zeldovich and Novikov (1967)
for the matter distribution over a $n=3$
polytrope.
The term containing the Newtonian gravitational energy of equilibrium star
may be explicitely evaluated for arbitrary polytropic
equation of state (Landau and Lifshits, 1976).
\begin{equation}
  \label{ref34.12}
  \varepsilon_G=-{3\over 5-n}\,{GM^2\over R}.
\end{equation}
  For an adiabat with
$\gamma=\gamma_{\rm {ad}}$, we have $E=n{P\over \rho}$ and for a star
in equilibrium takes place
\begin{equation}
\label{ref34.13}
  \varepsilon_{\rm i}=-{n\over 3}\,\varepsilon_G={n\over 5-n}\,{GM^2\over R},\quad
  \varepsilon_{\rm  N}=\varepsilon_{\rm i}+\varepsilon_G={n-3\over 5-n}\,{GM^2\over 3},
\end{equation}
 Where $\varepsilon_{\rm  N}$ is the total energy of a newtonian star. The
total energy of a stable star is negative, therfore the stability
requires that $n<3$, $\gamma>4/3$. The radius of a polytrope is, using
(\ref{ref34.2}), (\ref{ref34.5}) and (\ref{ref34.6})
\begin{equation}
\label{ref34.14}
  R=\alpha\xi_1=\left[{(n+1)\over 4\pi G}K\right]^{1/2}
  \rho_c^{{1-n \over 2n}}\xi_1
  =\left(\xi_1^3\over 4\pi M_n\right)^{1/3}M^{1/3}\rho_c^{-1/3}=
  {M^{1/3}\rho_c^{-1/3}\over 0.426}.
\end{equation}
 Here are used the values for a polytrope of $n=3$:
 $\xi_1=6.89685$, $M_3=2.01824$.
The ratio of $\rho_c$ to the average density $\overline\rho\left(M=
{4\pi\over 3}\overline\rho R^3\right)$ is
 $ {\rho_c\over \overline\rho}={4\pi\over 3}\,{1\over (0.426)^3}=54.18$.
 From (\ref{ref34.14}), (\ref{ref34.12}) and (\ref{ref34.7}) we have
\begin{equation}
  \label{ref34.16}
  \varepsilon_G=-0.639\,GM^{5/3}\,\rho_c^{1/3},\quad
  \varepsilon_{GR}=-0.918\,{GM^{7/3}\over c^2}\,\rho_c^{2/3}.
\end{equation}
 Only one parameter, $\rho_c^{1/3}$ or $R$, varies with homologous
variations:
\begin{equation}
  \label{ref34.17}
  \rho=\rho_c\varphi(\nu),\quad \nu=\frac{m}{M}, \quad
  \varphi(\nu)~~\hbox{is an invariant function}
\end{equation}
 and hence, the energy variations reduce to ordinary derivatives.
Using (\ref{ref34.16})--(\ref{ref34.17}) and taking the entropy to be constant at variations,
we get from (\ref{ref34.7}) the equilibrium condition
\begin{equation}
\label{ref34.18}
  \p{\varepsilon}{\rho_c^{1/3}}=3\rho_c^{4/3}\int_0^M P
  {dm\over \varphi(\nu)}\,-0.639\,GM^{5/3}
  -1.84\,{G^2M^{7/3}\over c^2}\,{\rho_c^{1/3}}=0.
\end{equation}
 The second derivative of the energy at $S={\rm const}$ turns into
zero on the boundary of stability:
\begin{equation}
\label{ref34.19}
  \p{^2\varepsilon}{\left(\rho_c^{1/3}\right)^2}=9\rho_c^{-5/3}\int_0^M
  \left(\gamma-{4\over 3}\right)P{dm\over \varphi(\nu)}
  -1.84\,{G^2M^{7/3}\over c^2}=0.
\end{equation}
 We use here the thermodynamic relations
$$    \p{E}{\rho_c^{1/3}}=3\rho_c^{2/3}\p\rho{\rho_c}\left(\p E\rho\right)_S
    =3\rho_c^{2/3}\varphi(\nu){P\over \rho^2}
    =3\,{P\over \varphi(\nu)}\,\rho_c^{-4/3},
$$
\begin{equation}
\label{ref34.20}
  \p{P}{\rho_c^{1/3}}=3\rho_c^{2/3}\left(\p P\rho\right)_S
  \varphi(\nu)
  =3\gamma P\rho_c^{-1/3},\quad
  \gamma\equiv\gamma_1=\plni P\rho S.
\end{equation}
 Equations (\ref{ref34.18}) and (\ref{ref34.19}) have been obtained
 by Bisnovatyi-Kogan (1966) and used
for determining the boundary of stability for various stellar models.

\section{Supermassive stars with a hot dark matter}

For $M>10^4M_{\odot}$ the main reason of instability are
GR effects. The entropy of such supermassive stars in critical state
is so large that the pressure is determined mainly by the radiation with
a small admixture of plasma, important for stability, but giving a very
short time until the onset of instability.
A common way to overcome this instability is to consider rotating
superstars, what may postpone the moment of collapse to $3 \times 10^4$
years for solid body rotation with angular momentum and mass losses
(Bisnovatyi-Kogan, Zeldovich and Novikov, 1967), and much longer for a
differentially rotating star evolving with almost constant angular
momentum (Fowler, 1966; Bisnovatyi-Kogan and Ruzmaikin, 1973).
Formation of supermassive stars
on early stages of the Universe expansion, their loss of stability
with subsequent collapse or explosion (Bisnovatyi-Kogan, 1968; Fricke, 1973;
Fuller et al, 1986) could be important for early formation
of heavy elements, observed in spectra of the most distant objects with
red shift $\sim 5$, creation of perturbations for large scale structure
formation, influence on small scale fluctuations of microwave background
radiation (Peebles, 1987; Cen et al, 1993). A necessity of a presence of a
dark matter in modern cosmological models makes it important to include
it into stability analysis of supermassive stars. This was done by
McLaughlin and Fuller (1996), who dealed with nonrotating superstars.
The same problem for rotating superstars,
using energetic method, was solved by Bisnovatyi-Kogan (1998).
The rotational effects occure to be more important
for realistic choice of parameters.

\subsection{Stability analysis}

In supermassive stars with equation of state $P=P_r+P_g=\frac{aT^4}{3}+
\rho {\cal R}T$, 

\noindent
( where $\cal R$ is a gas constant, $a$ is a constant of a
radiative energy density ) there is $P_r \gg P_g$ due to high entropy of
such stars.  Besides, such stars are fully convective and entropy is uniform
over them, so the spatial structure is well described by a polytropic
distribution, corresponding to $\gamma=4/3$.  The influence of a hot dark
matter, which density does not change during perturbations, should be taken by
account of a newtonian gravitational energy of the star in the dark matter
potential, because GR effects of a dark matter are of a higher order of
magnitude (McLaughlin, Fuller, 1996).  For radiation dominated plasma there is
a following expression for the adiabatic index, determining the stability to a
collapse

\begin{equation}
\label{ref1}
 \gamma=\left(\frac{\partial\log P}{\partial \log\rho}\right)_S \approx
 \frac{4}{3}
 \left(1+\frac{\cal R}{2S}\right)=\frac{4}{3}+\frac{\beta}{6},
 \end{equation}
 where $\beta=\frac{P_g}{P}= \frac{4{\cal R}}{S}$.
 In the radiationaly dominated supermassive
 star there is a unique connection between its mass $M$ and entropy per
 unit mass $S$ (Zeldovich and Novikov, 1967)

\begin{equation}
\label{ref2}
 M=4.44 \left(\frac{a}{3G}\right)^{3/2}\left(\frac{3S}{4a}\right)^2,
 \end{equation}
where $a$ is a constant of the radiation energy density, and numerical
coefficient is related to the polytropic density distribution with $\gamma=
4/3$. At the point of a loss of stability the critical value of an
average adiabatic index
$<\gamma>$ in selfgravitating nonrotating star with account of post-newtonian
corrections is determined by a relation (Zeldovich and Novikov, 1967)

\begin{equation}
\label{ref3}
<\gamma>_{crs}=\frac{4}{3}+ \delta_{GR}=
\frac{4}{3}+ \frac{2}{3} \frac{\varepsilon_{GR}}
{\varepsilon_G} \approx \frac{4}{3}+ 0.99 \frac{GM^{2/3} \rho_c^{1/3}}{c^2}.
\end{equation}
Here averaging is done according to (\ref{ref34.19}).
From comparison between (\ref{ref1}) and (\ref{ref3})
we get a well known relation for a critical central density of a
supermassive star stabilized by plasma

\begin{equation}
\label{ref4}
\rho_c=0.10\frac{{\cal R}^3 c^6}{G^{21/4} a^{3/4}} M^{-7/2} \approx
1.8 \times 10^{18} \left(\frac{M_{\odot}}{M}\right)^{7/2} {\rm g/cm}^3.
\end{equation}
Here and below we consider for simplicity a pure hydrogen plasma.
Newtonian energy of a superstar $\varepsilon_{nd}$ in the gravitational field of uniformly
distributed dark matter with a density $\rho_d$ is written as

\begin{equation}
\label{ref5}
\varepsilon_{nd}=\int_0^M \varphi_d dm.
\end{equation}
The gravitational potential of a uniform dark
matter $\varphi_d$ is written as

\begin{equation}
\label{ref6}
\varphi_d=\frac{2 \pi}{3}G\rho_d r^2-\frac{3}{2}\frac{GM_d}{R_d},
\end{equation}
where $R_d$ is much larger then stellar radius $R$, and $M_d$ is a total mass
of the dark matter halo.
Stability does not depend on normalization of the gravitational potential
so we shall omit the constant value in (\ref{ref5}).
It follows from (\ref{ref5}) and (\ref{ref6}) that during variations
$\varepsilon_{nd} \sim \rho_c^{-2/3}$, while
$\varepsilon_{GR} \sim \rho_c^{2/3}$  and
$\varepsilon_{G} \sim \rho_c^{1/3}$ (Zeldovich and Novikov, 1967).
Critical value of the adiabatic index is
obtained from (\ref{ref34.19}), and for nonrotating superstar
in presence of a dark matter the critical value of an average adiabatic
index $<\gamma>_{crnrot}$ is determined by

\begin{equation}
\label{ref7}
<\gamma>_{crnrot}= <\gamma>_{crs} + \delta_{dm}=
\frac{4}{3}+ \frac{2}{3} \frac{\varepsilon_{GR}}
{\varepsilon_G} - 2 \frac{\varepsilon_{nd}}{\vert\varepsilon_G\vert}.
\end{equation}
The relation for a critical density
in presence of a dark matter is obtained by comparison of
(\ref{ref1}) and (\ref{ref7}). We get

 \begin{equation}
 \label{ref8}
 0.99\frac{GM^{2/3} \rho_c^{1/3}}{c^2}=
 4.1\frac{\rho_d}{\rho_c}+\frac{\beta}{6}=
 4.1\frac{\rho_d}{\rho_c}+
 \frac{\cal R}{2a}\left(\frac{4.44}{M}\right)^{1/2}
 \left(\frac{a}{3G}\right)^{3/4}.
 \end{equation}
 Here relations (\ref{ref34.14}) is used for $R$, and
 $\int_0^R \rho r^4 dr=6.95 \times 10^{-4} \rho_c R^5$, based on Emden
 polytropic distribution with $\gamma=4/3$, was used (see Bisnovatyi-Kogan,
 1989). Relation (\ref{ref8}) is reduced to

\begin{equation}
\label{ref9}
2.8 \times 10^{-3}\left(\frac{M}{M_6}\right)^{2/3} \rho_c^{4/3}=
3.5 \times 10^{-4}\left(\frac{M_6}{M}\right)^{1/2} \rho_c+\rho_d.
\end{equation}
 Solution of (\ref{ref9}) is presented in Fig.1.

\subsection{Stability of rotating stars}
Consider a rigid rotation, when its energy is a small correction to
the energy of radiation and the energetic method is a good approach.
When losses of an angular momentum during evolution are negligible we
distinguish between rapidly rotating (RR) and slowly rotating (SR)
superstars. In RR case a superstar reaches the state of rotational
equatorial breaking before loosing its dynamical instability, and
in SR case instability comes first. If a superstar has an angular momentum
$J$, then its rotational energy $\varepsilon_{rot} \approx 1.25
J^2\rho_c^{2/3}M^{-5/3}$, and a ratio $\varepsilon_{rot}/\varepsilon_{GR}$
remains constant during evolution
(Bisnovatyi-Kogan, Zeldovich,
Novikov, 1967).
In presence of rotation and dark matter the critical
value of the adiabatic index $<\gamma>_{crrot}$ is written as

\begin{equation}
\label{ref10}
<\gamma>_{crrot}=
\frac{4}{3}+ \frac{2}{3} \frac{\vert\varepsilon_{GR}\vert-\varepsilon_{rot} }
{\vert\varepsilon_G\vert} -
2 \frac{\varepsilon_{nd}}{\vert\varepsilon_G\vert},
\end{equation}
and the relations for determination of a critical central density, instead
of (\ref{ref9}), is written as

\begin{equation}
\label{ref11}
2.8 \times 10^{-3}\left(\frac{M}{M_6}\right)^{2/3} \rho_c^{4/3}
\left(1-\frac{\varepsilon_{rot}}{\vert\varepsilon_{GR}\vert}\right)=
3.5 \times 10^{-4}\left(\frac{M_6}{M}\right)^{1/2} \rho_c+\rho_d.
\end{equation}
As follows from (\ref{ref11}), a superstar does dot loose its stability
when $\varepsilon_{rot} > \vert\varepsilon_{GR}\vert $. This qualitative
result, obtained in the post-newtonian approximation, remains to be valid
in a strong gravitational field and reflects
a presence of a limiting specific angular momentum $a_{lim}=GM/c$,
so that a black holes with a Kerr metric may exist only at $a< a_{lim}$
(Misner, Thorne, Wheeler, 1973).

A RR superstar in a course of the evolution reaches instead a limit of
a rotational instability, and equatorial mass shedding begins, leading to
a loss of an angular momentum. Such star will loose the stability when the
anuglar momentum will become less then the limiting value. The stage of a
mass loss was examined by
(Bisnovatyi-Kogan, Zeldovich, Novikov, 1967), where it was shown that
this stage may last about 10 times longer, then a maximum evolution time
to approach the rotational instability point.
RR star reaches the stage of a rotational instability at different central
densities, depending on $J$, but the ratio of rotatonal and Newtonan
gravitational energy on the mass-shedding curve remains constant
(Bisnovatyi-Kogan, Zeldovich, Novikov, 1967)
\footnote{Note, that
in presence of a dark matter halo mass outflow begins at rotational
energy approximately $(1+54\rho_d/\rho_c)$ times larger,
then in (\ref{ref12}) due to additional
gravity of a dark matter. This correction is small for a considered halo
density on the mass-shedding curve at $J>J_0$ (see below), and is
neglected in this section.}

\begin{equation}
\label{ref12}
\varepsilon_{rot}=0.00725\vert\varepsilon_{G}\vert.
\end{equation}
The energy of a rotating supertar in equilibrium in presence of a
hot dark matter may be written as

\begin{equation}
\label{ref13}
\varepsilon_{eq}=
-\varepsilon_{gas}+\vert\varepsilon_{GR}\vert
-\varepsilon_{rot}+3\varepsilon_{nd}.
\end{equation}
In the main term for a superstar in equilibrium  a relation is valid

\begin{equation}
\label{ref14}
\varepsilon_{gas}=\frac{\beta}{2}\vert\varepsilon_{G}\vert.
\end{equation}
Taking into account (\ref{ref12}), (\ref{ref14}), we get an expression for
an equilibrium energy along the mass-shedding curve (with variable $J$)

\begin{equation}
\label{ref15}
\varepsilon_{eq}=
-\left(0.00725+\frac{\beta}{2}\right)\vert\varepsilon_G\vert+
\vert\varepsilon_{GR}\vert+3\varepsilon_{nd}.
\end{equation}
The curve $\varepsilon_{eq}(\rho_c)$ has a minimum at the central density,
determined by a relation

\begin{equation}
\label{ref16}
2.8 \times 10^{-3}\left(\frac{M}{M_6}\right)^{2/3} \rho_c^{4/3} =
3.5 \times 10^{-4}\left(\frac{M_6}{M}\right)^{1/2} \rho_c+
5.9 \times 10^{-4}\rho_c+\rho_d.
\end{equation}
From comparison (\ref{ref16}) and (\ref{ref11}) with account of (\ref{ref12})
it is clear, that
dynamical instability cannot occure in the minimum of the mass-shedding
curve, and after crossing it the evolution proceeds with a substantial
mass and angular momentum losses. Central density of the superstar
in the minimum of the mass-
shedding curve (\ref{ref15})
with and without dark matter are represented in the fig.1.

Let us find a parameters of a superstar, at which its critical state
is situated on the mass-shedding curve. These parameters satisfy
sumultanously the relations (\ref{ref11}) and (\ref{ref12}),
what leads to the equation for determination of a central density
in the form

\begin{equation}
\label{ref17}
2.8 \times 10^{-3}\left(\frac{M}{M_6}\right)^{2/3} \rho_c^{4/3} =
3.5 \times 10^{-4}\left(\frac{M_6}{M}\right)^{1/2} \rho_c+
12 \times 10^{-4}\rho_c+\rho_d.
\end{equation}
The relation (\ref{ref12}) is used for determination of
an angular momentum of the superstar $J=J_0$ with $\rho_c$ from
(\ref{ref16}) and $J=J_1<J_0$ with $\rho_c$ from (\ref{ref17}).
Solution of this equation is also given in fig.1. As may be seen from fig.1,
the stabilizing effect of rotation on the mass-shedding curve
at $J=J_1$ is more
important, then stabilization by a hot dark matter, which influence
decreases at increasing of a central density. With
account of a longer state
of the evolution with mass loss until reaching the point of the loss
of stability at larger $J$,
this conclusion becomes even stronger.

\begin{figure}
\vspace{7cm}
\caption
{The correction terms  $\delta_{GR}$ and $\delta_{dm} + \delta_{GR}$,
the quantities $\beta$/6 (line c),
$\beta$/6+$(\varepsilon_{rot}/\varepsilon_N)_{sh}$/3 (line b), and
2$\beta$/6+
$(\varepsilon_{rot}/\varepsilon_N)_{sh}$/3 (line a),
 as functions of the central density of a
supermassive star with $M=10^6 M_{\odot}$, and dark matter density
of $10^{-5}$ g/cm$^3$. The instability points for nonrotating star
occure at intersection of the
correction term curves with the line c. Mass shedding in the stable
star with angular momentum $J_0$ (see text) occures at intersection of
correction term curves with the line b, and critical point on the
mass-shedding curve is determined by a corresponding intersection with
the line a.}
\label{fig1}
\end{figure}

\section{Collapse and explosions of supermassive stars}

To study dynamical processes by the energetic method we use, instead of
the energy variation, the energy conservation law in the form

\begin{equation}
\label{ref2.1}
d(\varepsilon_{in}+\varepsilon_{G}+\varepsilon_{GR}+\varepsilon_k)=
\int_0^M dQ dm=dS\int_0^M T dm.
\end{equation}
Here $\varepsilon_k=\frac{1}{2}\int_0^M v^2 dm $ is the kinetic energy of the
superstar. Using thermodynamic relation we get

$$d\varepsilon_{in}=d\int_0^M E(\rho,S) dm
=\int_0^M \left(\frac{\partial E}{\partial \rho}\right)_S d\rho\, dm
+\int_0^M \left(\frac{\partial E}{\partial S}\right)_{\rho}dS\, dm
$$
\begin{equation}
\label{ref2.2}
= d\rho_c\int_0^M \frac{P}{\rho^2}\frac{d\rho}{d\rho_c}\, dm
 +dS\int_0^M T dm.
\end{equation}
 From the mass conservation law, in presence of homologycal motion with
 the fixed density distribution in space (\ref{ref34.17}), we obtain a
 space velocity distribution. Considering $r$ as a Lagrangian
 radius, corresponding to the mass $m \le M$, $r(M)=R$, we may write

$$ m=A(\xi) \rho_c r^3, \quad M=A_0 \rho_c R^3,\quad A(\xi_1)
    =A_0=\frac{4\pi}{\xi_1^3}M_n,$$
\begin{equation}
\label{ref2.3}
\frac{d\rho_c}{dt}+3\frac{\rho_c}{r}\frac{dr}{dt}=0, \quad
\frac{d\rho_c}{dt}+3\frac{\rho_c}{R}\frac{dR}{dt}=0.
\end{equation}
Using definitions of the velocity in Lagrangian coordinates, we get
its homological space distribution

\begin{equation}
\label{ref2.4}
v=\frac{dr}{dt}, \quad v_R=\frac{dR}{dt}, \quad
v=v_R\frac{r}{R}.
\end{equation}
With account of the equality $\int_0^1(r^2/R^2)\varphi(\nu)d\nu=0.108$,
we get a dynamical equation in the form

\begin{equation}
\label{ref2.5}
0.597\frac{d^2(\rho_c^{-1/3})}{dt^2}+0.639\,G\rho_c^{2/3}
+1.84\,{G^2M^{2/3}\over c^2}\,{\rho_c}
\end{equation}
$$-3M^{-2/3}\rho_c^{-2/3}\int_0^1 P {d\nu\over \varphi(\nu)}\,=0.$$
On dynamical stages of the superstar evolution, after its loss of stability
the radiative transfer losses $(-Q_r)$ are
less important, then neutrino losses $(-Q_{\nu})$ and
energy production in nuclear reactions $Q_n$.
The equation determining entropy  changes is
averaged over the star with a weight function $(T^4 \sim E\rho)$ for
radiation dominated superstar, which entropy is taken homogenous

\begin{equation}
\label{ref2.6}
\frac{dS}{dt} \int_0^1 \rho T^5 d\nu=\,<T^4\,Q_n>-<T^4\,Q_{\nu}>-<T^4\,Q_r>.
\end{equation}
Nuclear reactions of $pp$ and $CNO$ hydrogen burning and $3\alpha$ helium
burning have been included in the calculations of Bisnovatyi-Kogan (1968).
The equaions (\ref{ref2.5}),(\ref{ref2.6}),
together with equations for averaged composition of hydrogen $X$, helium
$Y$, and equation of state with account of $e^+\,e^-$ pairs creation and
relativistic relations for electrons, had been solved numerically.
Primodial chemical composition with only hydrogen and helium was
taken as initial condition. In the process of contraction after a loss
of stability, $3\alpha$ reaction produces $^{12}C$, which initiates a
$CNO$ hydrogen burning, $pp$ reaction remaining always unimportant.
It was obtained that in stars with
$M< 1.5\times 10^5 M_{\odot}$ collapse is reversed,
and they explode, enriching the intergalactic and interstellar gas with
heavy elements. Such explosions could happen on stages, preceding
the epoch of a galaxy formation. That is one of the way to explain
high metallicity in the distant quasars and intergalactic gas in
galaxy clusters. Similar calculations made for rotating superstars,
and for normal (solar) composition shift the boundary between collapsing and
exploding superstars to higher masses (Fricke,1973).

\section{Formulation of Galerkin method}

The Galerkin method allows to find an approximate solution of differential
equations.
In this method the solution of the partial or ordinary differential
equation is represented in the form (Fletcher, 1984)

\begin{equation}
\label{galerkin}
u=u_0+\sum_{i=1}^N \alpha_i \varphi_i \mbox{,}
\end{equation}
where $\varphi_i$ are known analytical functions, and
coefficients (or functions)
$\alpha_i$ are to be determined. Functions $\varphi_i$ are assumed to be
not necessary orthogonal. When substituted to the original differential
equation the coefficients $\alpha_i$ will satisfy a set of algebraic
or ordinary differential equations, obtained by
minimizing the corresponding functional.

Consider how the
Galerkin method can be applied to the problem of the
relativistic star collapse.
The functional which should be minimized, is a total stellar energy. It
contains a kinetic energy (\ref{ref2.1}), which in PN is written explicitly.
In GR the total energy cannot be splitted into different components, and the
collapsing body should be described by means of a stress energy tensor
$T_{ij}$. Equating to zero the variation of the total energy gives
dynamical equations of the collapse. In the process of contraction the
collapse accelerates and finally approaches a free fall which is described
by a Tolman solution (Landau and Lifshits, 1962). The approximate method
should be formulated in such a way, that its solution approaches the
Tolman one at final stages.

We may use the Galerkin method to approximate the system of
equations in partial derivatives by the system of ordinary differential
equations.
The density, as a function of a lagrangian coordinate $a$ and time $t$,
$\rho(a,t)$ is
represented as a sum of $\alpha_i(t)\varphi_i(a)$. Since $\varphi(a)$ is
invariant under the lagrangian transformations, the partial time derivative
of this product
will transform to $\frac{d\alpha_i}{dt}\varphi_i(a)$.
Thus,
the Galerkin method allows to separate variables $t$, and $a$.
The same separation has to be done for a velocity function $v(a,t)$.

\subsection{Stability conditions in PN}

In PN approximation the density in Galerkin method is written
as a following sum

\begin{equation}
\label{6}
\rho=\sum_{i=1}^N\alpha_i(t) \varphi_i(a)
\qquad\mbox{,}\qquad\mbox{where}\qquad
\rho_c=\sum_{i}^N\alpha_i(t)\varphi(0)
\end{equation}
As for a function
$\varphi_0(a)$, it is convenient to take corresponding Emden profile for
one of polytropic indices. For other functions we may chose
$\varphi_k=cos\frac{1+2K}{2}\pi a$, which have increasing number of nodes.

Then satisfaction
of the boundary conditions:

\begin{equation}
\label{boundary}
\varphi_i (A)=0 \mbox{,}\quad A=a(R); \qquad \varphi_i (0)=1
\end{equation}
will be provided. The minimization of the energy functional (\ref{ref34.7})
for finding an equilibrium model is reduced to zero partial derivatives

\begin{equation}
\label{ref3.1}
\frac{\partial\varepsilon}{\partial\alpha_i}=0,
\end{equation}
leading in the static case of constant $\alpha_i$ to a set of $N$ algebraic
equations for finding equilibrium $\alpha_i^{eq}$. Stability of a model
is found from an evaluation of the second variation $\delta^2\varepsilon$,
which in the energetic method is reduced to the algebraic equation
(\ref{ref34.19}). In the Galerkin method with several scaling functions
$\varphi_i(a)$, the second variation $\delta^2\varepsilon$ is represented
by a quadratic form

\begin{equation}
\label{ref3.2}
\delta^2\varepsilon
=\sum_{i,k}^N \frac{\partial^2\varepsilon}{\partial\alpha_i\partial\alpha_k}
\delta\alpha_i\delta\alpha_k,
\end{equation}
In the energetic method the condition of stability reduces to one equation
$\partial^2\varepsilon/\partial\alpha^2>0$, given in (\ref{ref34.19}).
In the Galerkin method the stability is related to positive definiteness
of the quadratic form (\ref{ref3.2}), what is provided (Smirnov, 1958) by
the positiveness of the determinant

\begin{equation}
\label{ref3.3}
\Vert\frac{\partial^2\varepsilon}{\partial\alpha_i\partial\alpha_k}\Vert
>0,
\end{equation}
and all its main minors. Remind, that main minors are determinants, obtaind
from the main determinant (\ref{ref3.3}) after eliminating
lines and columns, intersecting on the main diagonal at $i=k$. For two
functions in (\ref{6}) the positiveness of the main determinant
(\ref{ref3.3}), and two partial derivatives
$\partial^2\varepsilon/\partial\alpha_1^2>0$ and
$\partial^2\varepsilon/\partial\alpha_2^2>0$ are enough for stellar
stability. Loss of stability happens close before the point where
the main determinant, or one of its main minors becomes zero.

In approximate presentation of the trial function in the Galerkin method,
the minimal value of the second energy variation is larger, then its
value for a real trial function. So zero values of the determinant
(\ref{ref3.3}), or one of its main minors, guarantees
the onset of instability.
Their positiveness is not an exact guarantee of the stability, but comparison
of the energetic method with an exact stability analysis shows a
good presicion of this approximate approach in most realistic cases
(Bisnovatyi-Kogan, 1989). Energetic method corresponds to a homologeous
trial function for displacement $\delta r \sim r$. In the Galerkin
method the trial function may be determined with a better precision.
In fact, the coefficients $\delta \alpha_i$ for the trial function of
a density

\begin{equation}
\label{ref3.4}
\delta\rho=\sum_i^N\delta\alpha_i\varphi_i(a)
\end{equation}
are determined as an eigenvector of a set of uniform linear equations

\begin{equation}
\label{ref3.5}
\frac{\partial^2\varepsilon}{\partial\alpha_i\partial\alpha_k}
\delta\alpha_k=\lambda_p\delta\alpha_i.
\end{equation}
The eigenvector $\delta\alpha_i^e$ is used for obtaining an approximate
eigenfunction (\ref{ref3.4}), and eigenvalues $\lambda_p$ are related
to the square eigenfrequencies of the stellar model. The positive
definiteness of the quadratic form (\ref{ref3.2}) coincides with the
positiveness of all eigenvalues $\lambda_p$. So, Galerkin method
permits to investigate a stability by finding
approximate eigenvalues and eigenfunctions of a linear set of
algebraic equations, instead of finding the same values from a second order
differential equation exactly.

\subsection{Relations for total energy and barion number in GR}

In GR a barion number density $n$ is used in (\ref{6}) instead of the
mass density $\rho$, which in this case is not presenting itself a
conserved value.
To obtain GR expression for the total energy of a spherically symmetric
radially moving body take a
stress-energy tensor for the perfect gas

\begin{equation}
\label{Tij}
T_{ij}=(P+n\epsilon)u_i u_j-P g_{ij}\mbox{,}
\end{equation}
where $\epsilon$ is an
internal energy per baryon, $u_i$  is a four velocity of the matter.
Inside the spherically symmetric star the Schwarzschild type metric is
used

\begin{equation}
\label{geuler}
g_{ij}=e^{\phi}c^2 dt^2-e^\lambda dr^2-r^2 d\Omega^2\mbox{,}
\end{equation}
where $d\Omega^2= d\theta^2+\sin^2\theta d\varphi^2$, and
$\lambda$ is expressed as a functional of $T^0_0$ (Landau and Lifshits, 1962)

\begin{equation}
\label{e}
e^\lambda=(1-\frac{2Ge}{c^4r})^{-1}, \quad
e=4\pi\int\limits_0^r T_0 ^0 r^2\,dr.
\end{equation}
To avoid confusion, we determine $r$ and
$t$ to be independent Euler variables and $v$ to be a velocity field
of the medium.
A number of baryons $a$ inside a Lagrangian
radius $r(a)$ is considered as a Lagrangian independent coordinate

\begin{equation}
\label{a}
a=4\pi\int_0^r \frac{n r^2}{\sqrt{1-\frac{2Ge}{c^4r}}} \,dr.
\end{equation}
Then, the world line of each shell is $r(t,a)$. There
exist a set of transformations of the Euler type metric ~(\ref{geuler}) to
its Lagrangian representation:

\begin{equation}
\label{glagr}
g_{ij}=e^{2\nu}c^2 d\tau^2-e^{2\Lambda} da^2-r^2 d\Omega^2\mbox{,}
\end{equation}
where $\tau$ scales time of the co-moving observer.
We determine a physical velocity of the
shell as a displacement $dl$, measured by the Euler rest observer,
with respect to Euler time, measured by the same observer:

\begin{equation}
\label{velocity}
v=\frac{e^{-\phi/2}}{\sqrt{1-\frac{2Ge}{c^4r}}}
\left(\frac{\partial r}{\partial t}\right)_a.
\end{equation}
Where $\phi$ to be determined from (\ref{geuler}).
In a case of a test particle it gives a well known result for
the free falling particle in a Schwarzschild metric

\begin{equation}
\label{v1}
v=(1-\frac{2GM}{c^2r})^{-1}\frac{dr}{dt} =(1-\frac{2GM}{c^2r})^{-1} \dot r,
\quad \dot r=\frac{dr}{dt}=v^r.
\end{equation}
Covariant components $v_{\alpha}$ and $v^2$
are given by the following relations
(Landau and Lifshits, 1962)

\begin{equation}
\label{v2}
v_\alpha=\gamma_{\alpha\beta}v^\beta\qquad\mbox{,}\qquad v^2
=v_\alpha v^\alpha.
\end{equation}
In a case of a diagonal metric, the dimensionless 3 - metric tensor
is connected with the 4 - metric tensor as
$\gamma_{\alpha\beta}=-g_{\alpha\beta}$. The mass-energy functional for the
spherically symmetrical relativistic star is written as in (\ref{e})

\begin{equation}
\label{14}
E=e(R)=4\pi\int\limits_0^R T_0 ^0 r^2\,dr .
\end{equation}
Four velocity of the fluid can be expressed in terms
of the tree velocity (Landau and Lifshits, 1962)

\begin{equation}
\label{u}
u^\alpha=\frac{v^\alpha}{\sqrt{g_{00}}\sqrt{1-\frac{v^2}{c^2}}} , \qquad
u^0=\frac{1}{\sqrt{g_{00}} \sqrt{1-\frac{v^2}{c^2}}}.
\end{equation}
Then, $T^0_0$ component of the stress-energy tensor (\ref{Tij})
is written as

\begin{equation}
\label{T00}
T^0_0=\frac{n\epsilon+P \frac{v^2}{c^2}}{1-\frac{v^2}{c^2}}.
\end{equation}
From (\ref{e}),(\ref{a}) making use of ~(\ref{T00}) we get

\begin{equation}
\label{first}
\frac{\partial e}{\partial a}=\frac{n\epsilon+P \frac{v^2}{c^2}}
{1-\frac{v^2}{c^2}}\sqrt{1-\frac{2Ge}{c^4r}},
\end{equation}
\begin{equation}
\label{second}
\frac{\partial r}{\partial a}=\frac{1}{4\pi n r^2}
\sqrt{1-\frac{2Ge}{c^4r}}.
\end{equation}

\subsection{Velocity distribution over the star in GR}

Assuming the profile law for $n$, let us obtain the profile law for the
velocity. To separate variables $a$ and $t$ we need a relation between
the pertubations of the boundary
$\left(\partial r / \partial t\right)_{a=N}$,
and velocity of the inner shell
$\left(\partial r / \partial t\right)_a$.
Differentiating ~(\ref{second}) gives

\begin{equation}
\label{rdot}
\frac{\partial\dot r}{\partial a}=\frac{1}{8\pi n r^2}
\frac{-2G\dot e/c^4 r+
2Ge\dot r/c^4 r^2}{\sqrt{1-\frac{2Ge}{c^4r}}}
-\frac{\sqrt{1-\frac{2Ge}{c^4r}}}{2\pi n r^2}\left(\frac{\dot n}
{2n}+\frac{\dot r}{r}\right),
\end{equation}
what contains terms with $\dot e$. After differentiation,
equation ~(\ref{first})
contains a second order derivative $\ddot r$. Going on this iterative
process, we obtain an infinite system of differential equations:

\begin{eqnarray}
\frac{\partial \dot e}{\partial a}
=\frac{((\dot{n\epsilon})+\dot P \frac{v^2}{c^2} +2 P \frac{v\dot v}{c^2})
\sqrt{1-\frac{2Ge}{c^4r}}} {(1-\frac{v^2}{c^2})}+
2\frac{v\dot v}{c^2}\frac{(n\epsilon+P \frac{v^2}{c^2})
\sqrt{1-\frac{2Ge}{c^4r}}}{(1-\frac{v^2}{c^2})^2}\nonumber\\
+\frac{1}{2}\frac{(n\epsilon+P \frac{v^2}{c^2})
\left(-2G\dot e/c^4 r+2Ge\dot r/c^4 r^2\right)}
{(1-\frac{v^2}{c^2})\sqrt{1-\frac{2Ge}{c^4r}})},\label{edott}\\
\frac{\partial \ddot e}{\partial a}=...\,,\label{last}\\
\vdots ,\label{llast}
\end{eqnarray}
completed by a similar set of equations obtained  from differentiation of
(\ref{rdot}).
In PN the velocity distribution (\ref{ref2.4}) is found from the continuity
equation, and than is used in derivation of the equation of motion
(\ref{ref2.5}). In full GR description the infinite set of equations
(\ref{edott})-(\ref{llast}) should be used for obtaining the velocity
distribution, instead of (\ref{ref2.4}).
Such complication is connected with an influence of the 'kinetic' energy
on the geometry of the star interior.
After substituting of expansion on $n$, similar to (\ref{6}),
we obtain ordinary differential equations instead of the partial ones.
In practice the system ~(\ref{edott}) - ~(\ref{llast})
has to be
cutted off on some step. Final stages of the gravitational collapse are
close to a free fall of a dust spherical cloud.
Thus using $\dot e$ from the Tolman solution should be good for late stages
of the collapse. For the zero-pressure dust collapse without thermal
processes the energy inside a given  lagrangian shell does not change,
so in (\ref{rdot}) the relation $\dot e(a)= (\partial e/\partial t)_a=0$
could be approximately used, making unnesessary to solve the equations
(\ref{edott}) - (\ref{llast}) for obtaining the profile $v(a)$.

\subsection{Equations of motion}

The equation determining the change of the entropy of the star is written
similar to (\ref{ref2.6}). In that case the entropy distribution over
the star should not change its form in the process of the collapse, or
be uniform. The integration should be done over the lagrangian coordinate
$da=
\frac{4\pi n r^2 dr}{\sqrt{1-\frac{2Ge}{c^4r}}}$.
To derive equation of motion we should, like in (\ref{ref2.5}), to take
derivatives from all thermodynamic functions at constant entropy distribution,
as follows from (\ref{ref2.1}),(\ref{ref2.2}).

When only one term is is left in (\ref{6}), as in the energetic method, the
equation (\ref{edott}) is used also for derivation of the
equation of motion. In the case of several terms in (\ref{6}), the equation
(\ref{edott}) should be written for each coefficient
$\alpha_k$, taking constant other $\alpha_i, \quad i \neq k$. This the number
of equations will be equal to the number of
unknown functions $\alpha_i$. This procedure is equivalent to a minimization
of the energy functional over all coefficients $\alpha_i(t)$, or over all
values of $\alpha_i$ for the equilibrium solution.

It is important to stress, that when using (\ref{edott}) - (\ref{llast})
for derivation of the velocity profile, all coefficients $\alpha_i(t)$
must be differentiated together.

\section{Conclusion}

We suggest an approximate method for solving the problem of stellar
stability, and dynamical stages of relativistic stellar collapse,
using Galerkin method.
It should be mentioned, that this paper, together with the general review of
the contemporary state of art of the problem contains the description of the
essentially new results. The stability analysis of the rotating stars with the
presence of the surrounding hot dark matter background, as well as the
application of the energetic method together with the Galerkin method to the
problem of relativistic gravitational collapse are such new results.

In the case of stability investigation we
get a set of algebraic equations and search of the eigenvalues of
matrics, instead of solution of eigenvalue problem for differential
equation of a second order. In the case of collapse set of ordinary
differential equations should be solved, instead of partial ones.
It is evident, that solution of ordinary equations is much simpler,
what could permit to investigate a wide set of equation of states, stellar
masses, and to get different neutrino light curves during a black
hole formation.

Several examples of application of the simplest version of Galerkin
method, a well known energetic method, are presented. A simple solution
obtained by this method for complicated problems show explicitely the
advantage of such approximate approach.

\bigskip

{\bf Acknowledgements} \\
This work was partly supported by Russian
Basic Research Foundation
grant No. 96-02-16553 and grant of a Ministry of Science and Technology
1.2.6.5.


\begin{thebibliography}{99}

\bibitem{shapiro}
Baumgarte T., Shapiro S., Teukolsky S., 1995,
ApJ, 443, 717

\bibitem{shapiro1}
Baumgarte T., Janka T., Keil W., Shapiro S., Teukolsky S., 1996,
ApJ, 468, 823

\bibitem{bk66}
Bisnovatyi - Kogan G.S., 1966,
Azh, 43, 89

\bibitem{bigstars}
Bisnovatyi - Kogan G.S.,1968,
Azh, 45, 74

\bibitem{3}
 Bisnovatyi-Kogan,G.S. 1989, Physical Problems of the
 Theory of Stellar Evolution (Moscow, Nauka)

\bibitem{bk98}
Bisnovatyi - Kogan G.S.,1998,
ApJ (in press)

\bibitem{4}
Bisnovatyi-Kogan,G.S., Ruzmaikin, A.A., 1973,
Astron. Ap., {\bf 27}, 209

\bibitem{5}
Bisnovatyi-Kogan,G.S., Zel'dovich,Ya.B., Novikov,I.D. 1967,
Astron. Zh. {\bf 44} 525

\bibitem{6}
Cen,R., Ostriker,J., Peebles,P.J.E. 1993, ApJ, {\bf 415}, 423

\bibitem{ch}
Chandrasekhar S., 1939,
Stellar Structure,
Chicago

\bibitem{galerkin}
Fletcher C.A.J., 1984,
Computational Galerkin methods,
Springer-Verlag, New York Berlin Heidelberg Tokyo

\bibitem{8}
Fowler,W. 1966, ApJ. {\bf 144}, 191

\bibitem{9}
Fuller,G.M., Woosley,S.E., Weaver,T.A. 1986, ApJ, {\bf 307}, 675

\bibitem{10}
Fricke,K. 1973, ApJ, {\bf 183}, 941

\bibitem{wheeler}
Harrison B.K., Thorne K.S., Wakano M. Weeler J.A. 1965,
Gravitation theory and gravitational collapse
The University of Chicago Press

\bibitem{landau}
Landau L., Lifshitz E., 1962,
The classical theory of fields. Nauka, Moscow

\bibitem{llsp}
Landau L., Lifshitz E., 1976,
Statistical Physics vol.1. Nauka, Moscow.

\bibitem{12}
McLaughlin,G., Fuller,G. 1996, ApJ, {\bf 456}, 71

\bibitem{13}
Misner Ch.W., Thorne K.S., Wheeler J.A. 1973. Gravitation. W.H.Freeman
and Co. San Fransisco.

\bibitem{14a}
Peebles,P.J.E. 1987, ApJ, {\bf 313}, L73

\bibitem{sm58}
Smirnov V.I. (1958),
Kurs Vysshey Matematiki, vol. 3, part 1.
Fizmatgiz, Moscow.

\bibitem{15}
 Zel'dovich,Ya.B., Novikov,I.D. (1965), Uspekhi Fiz. Nauk {\bf  86} 447.


\bibitem{zn67}
Zeldovich Ya. B. and Novikov I.D., 1967,
Relativistic Astrophysics. Nauka, Moscow

\end{thebibliography}
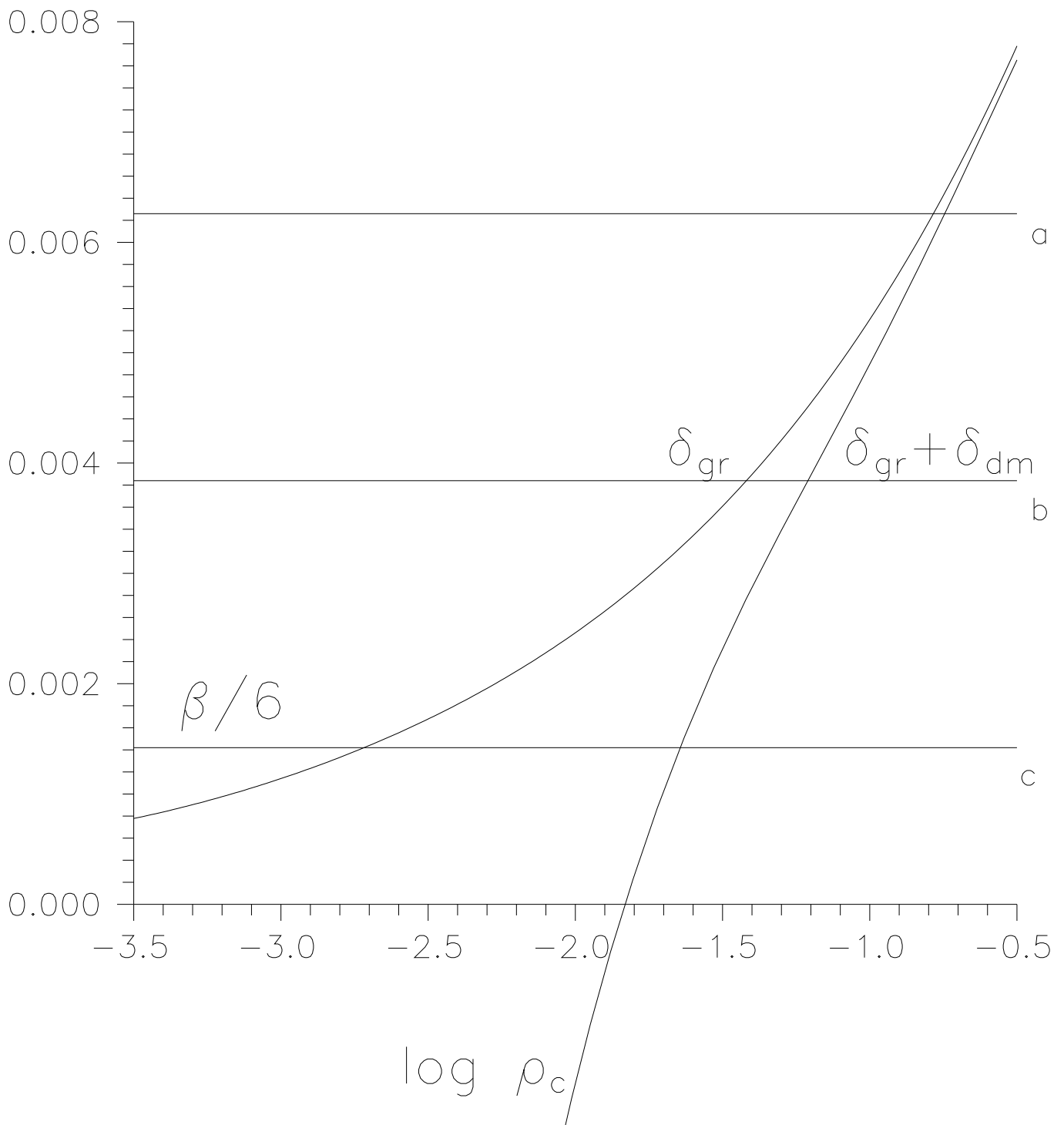
\end{document}